\documentclass[journal=ancac3,manuscript=article]{achemso}
\newif\ifshowcomments
\showcommentstrue

\usepackage{comment}
\usepackage[utf8]{inputenc}
\usepackage[T1]{fontenc}
\usepackage{amsmath,amssymb}
\usepackage{graphicx}
\usepackage{grffile}
\usepackage{xcolor}
\usepackage{booktabs}
\usepackage{array}
\usepackage[T1]{fontenc}
\usepackage[final]{microtype}
\usepackage{xurl}          
\usepackage{hyphenat}

\graphicspath{{Figures/}{figures/}}

\newcommand{\CuO}{Cu$_2$O}
\newcommand{\SiN}{SiN$_x$}
\newcommand{\um}{$\mu$m}

\newcommand{\placeholderfigure}[1]{%
  \fbox{\parbox[c][0.24\textheight][c]{0.88\linewidth}{\centering\small Placeholder for figure file:\\[0.3em] {\ttfamily\detokenize{#1}}\\[0.5em]Upload the figure to Overleaf using this file name, or replace the file name in the source.}}%
}
\newcommand{\safeincludegraphics}[2][]{%
  \IfFileExists{#2}{\includegraphics[#1]{#2}}{\placeholderfigure{#2}}%
}

\usepackage[
    colorlinks=true,
    linkcolor=blue,      
    citecolor=green,     
    filecolor=magenta,   
    urlcolor=cyan        
]{hyperref}

\title{Interfacing Silicon Nitride Nanophotonics with Cuprous Oxide Rydberg Excitons}

\author{Artem Kryvobok}
\altaffiliation{These authors contributed equally to this work.}
\affiliation{Elmore Family School of Electrical and Computer Engineering, Purdue University, West Lafayette, Indiana 47906, United States}

\author{Hamza Ather}
\altaffiliation{These authors contributed equally to this work.}
\affiliation{Department of Physics and Astronomy, Purdue University, West Lafayette, Indiana 47906, United States}

\author{Julio Herreros García}
\altaffiliation{These authors contributed equally to this work.}
\affiliation{Elmore Family School of Electrical and Computer Engineering, Purdue University, West Lafayette, Indiana 47906, United States}

\author{Ziqi Niu}
\affiliation{Elmore Family School of Electrical and Computer Engineering, Purdue University, West Lafayette, Indiana 47906, United States}

\author{Abhishek Bharatbhai Solanki}
\affiliation{Elmore Family School of Electrical and Computer Engineering, Purdue University, West Lafayette, Indiana 47906, United States}

\author{Vahagn Mkhitaryan}
\affiliation{Elmore Family School of Electrical and Computer Engineering, Purdue University, West Lafayette, Indiana 47906, United States}

\author{Benjamin Cerjan}
\affiliation{Department of Chemistry, Purdue University, West Lafayette, Indiana 47906, United States}

\author{Alexandra Boltasseva}
\affiliation{Elmore Family School of Electrical and Computer Engineering, Purdue University, West Lafayette, Indiana 47906, United States}

\author{Vladimir M. Shalaev}
\affiliation{Elmore Family School of Electrical and Computer Engineering, Purdue University, West Lafayette, Indiana 47906, United States}

\author{Hadiseh Alaeian}
\email{halaeian@purdue.edu}
\affiliation{Elmore Family School of Electrical and Computer Engineering, Purdue University, West Lafayette, Indiana 47906, United States}
\alsoaffiliation{Department of Physics and Astronomy, Purdue University, West Lafayette, Indiana 47906, United States}


\begin{document}

\begin{abstract}
Rydberg excitons in cuprous oxide (\CuO{}), the solid-state counterpart of Rydberg atoms, are hydrogen-like electron--hole quasi-particles with micron-sized wavefunctions and strong long-range interactions that promise optical nonlinearities at the few-photon level. Exploiting them in a waveguide quantum electrodynamics (WQED) setting can enable what atomic Rydberg platforms achieve only with ultrahigh vacuum and laser cooling/trapping---a guided optical mode interfaced with the Rydberg medium---while adding the scalability and advanced fabrication of integrated photonics. Here we demonstrate a monolithic, CMOS-compatible interface of \CuO{} Rydberg excitons with a \SiN{} nanophotonic waveguide. Under non-resonant excitation, excitonic emission from the \CuO{} microcrystal is coupled to the guided mode, routed along the waveguide, and collected from an output grating. The spectrum shows the phonon-assisted 1$s$ orthoexciton emission together with the low-lying 2$p$ and 3$p$ lines of the yellow Rydberg series. Spatially selective excitation excludes substrate background and direct scatter, and the enhanced linear polarization of the grating-collected signal identifies the TE-like guided mode as the collection channel. These results establish the complete material--device--measurement stack for a novel platform of nonlinear quantum nanophotonics based on Rydberg excitons.
\end{abstract}

\maketitle

\section{Introduction}~\label{sec:intro}
Optical nonlinearities at the few-photon limit are a central goal of quantum photonics,~\cite{Chang2014} yet conventional materials have far too weak optical nonlinearity at low photon numbers. Rydberg atoms circumvent this fundamental limit. Photons are mapped onto highly excited states with micron-scale wavefunctions and strong long-range interactions. These interactions induce energy shifts within a blockade volume, giving rise to a collective optical nonlinearity without the need for a cavity.~\cite{Peyronel2012, Dudin2012, Firstenberg2013, Gorniaczyk2014, Tiarks2014, Firstenberg2016} These compact nonlinear interaction regions are well suited to waveguide quantum electrodynamics, where a subwavelength guided mode enhances the interaction per photon within a well-defined, routable channel.~\cite{Chang2018RMP, Sheremet2023}
 
Atomic WQED has matured over the past decade: cold atoms have been trapped in the evanescent field of nanofibers,~\cite{Vetsch2010, Corzo2019} coupled to photonic crystal cavities and waveguides,~\cite{Thompson2013, Tiecke2014, Goban2014, Goban2015} loaded onto integrated microring circuits,~\cite{Chang2019Optica, Zhou2023PRL, Zhou2024PRX} and interfaced with chiral guided modes~\cite{Petersen2014, Lodahl2017}. Thermal vapors have been coupled to integrated waveguides, photonic crystal cavities, and microrings, reaching collective strong coupling without laser cooling.~\cite {Stern2013, Ritter2015, Ritter2018, Alaeian2020APB, Skljarow2020, Skljarow2022, Cheng2026} Atomic Rydberg exictation has been achieved in micrometer-sized vapor cells, hollow-core fibers, and at submicron distances from nanofibers.~\cite{Kubler2010, Epple2014, Rajasree2020} All these studies inherit the overhead of the atomic medium: vacuum, trapping and cooling, surface-induced loss of Rydberg states, and none offers control over the position of the nonlinear medium.
 
Solid-state ``artificial atoms'' remove this overhead: quantum dots,~\cite{Lodahl2015RMP, Arcari2014} color centers in diamond,~\cite{Sipahigil2016, Evans2018, Bhaskar2020, Wan2020} rare-earth ions,~\cite{Zhong2017, Dibos2018, Kindem2020} and emitters in two-dimensional materials\cite{Peyskens2019, ErrandoHerranz2021, Parto2022, Rivera2019, Qian2024, Liu2015, Kedziora2024} have been integrated with nanophotonic devices with near-unity coupling and deterministic positioning. What they lack is the Rydberg resource: their nonlinearity is that of a single two-level emitter, and where excited excitonic states are accessible, nanometer-scale Bohr radii limit the interaction range.~\cite{Gu2021, WS2PhC2025} Thus far, solid-state medium combining strong long-range Rydberg interactions with nanophotonic integration has been missing.
 
Cuprous oxide fills this gap. Its yellow exciton series extends to $n\approx25$--$30$, with micron-scale wavefunctions and an experimentally observed Rydberg blockade.~\cite{Kazimierczuk2014, Versteegh2021, Heckotter2021, Assmann2020} Giant nonlinearities are predicted for Rydberg excitons coupled to confined photonic modes.~\cite{Walther2018, WaltherPohl2020, Khazali2017} Experiments have shown Rydberg exciton--polaritons and their nonlinearities in microcavities,~\cite{Orfanakis2022, Makhonin2024} self-Kerr effects across the Rydberg series,~\cite{Morin2022} and microwave--optical coupling.~\cite{Gallagher2022, Pritchett2024} Crucially, Rydberg-grade \CuO{} no longer requires natural crystals: thermally oxidized copper on various substrates hosts Rydberg excitons up to $n\approx 9$,~\cite{Steinhauer2020, Lynch2021, DeLange2023} and is compatible with lithographic patterning,~\cite{Barua2025} which complements the top-down nano-structuring of bulk crystals.~\cite{Paul2024, Orfanakis2021}
 
The missing step is a nano-photonic device interface. 
Here we demonstrate this interface. Site-controlled \CuO{} microcrystals are integrated with a nitrogen-rich, low-autofluorescence \SiN{} waveguide on silicon using only CMOS-compatible processes. Under non-resonant excitation at cryogenic temperatures, emission from the \CuO{} island couples into the guided mode and is extracted at grating couplers tens of microns away, where the phonon-assisted 1$s$ orthoexciton emission and the 2$p$ and 3$p$ lines of the yellow Rydberg series are resolved. Spatially selective excitation rules out substrate luminescence and direct scatter, and the enhanced linear polarization of the grating-collected light identifies the TE-like guided mode as the collection channel. The result is the complete material--device--measurement stack of a solid-state Rydberg waveguide QED platform and the geometry for the resonant transmission, pump--probe, and photon-correlation experiments to follow.

\begin{figure}[t]
    \centering
    \safeincludegraphics[width=0.95\linewidth]{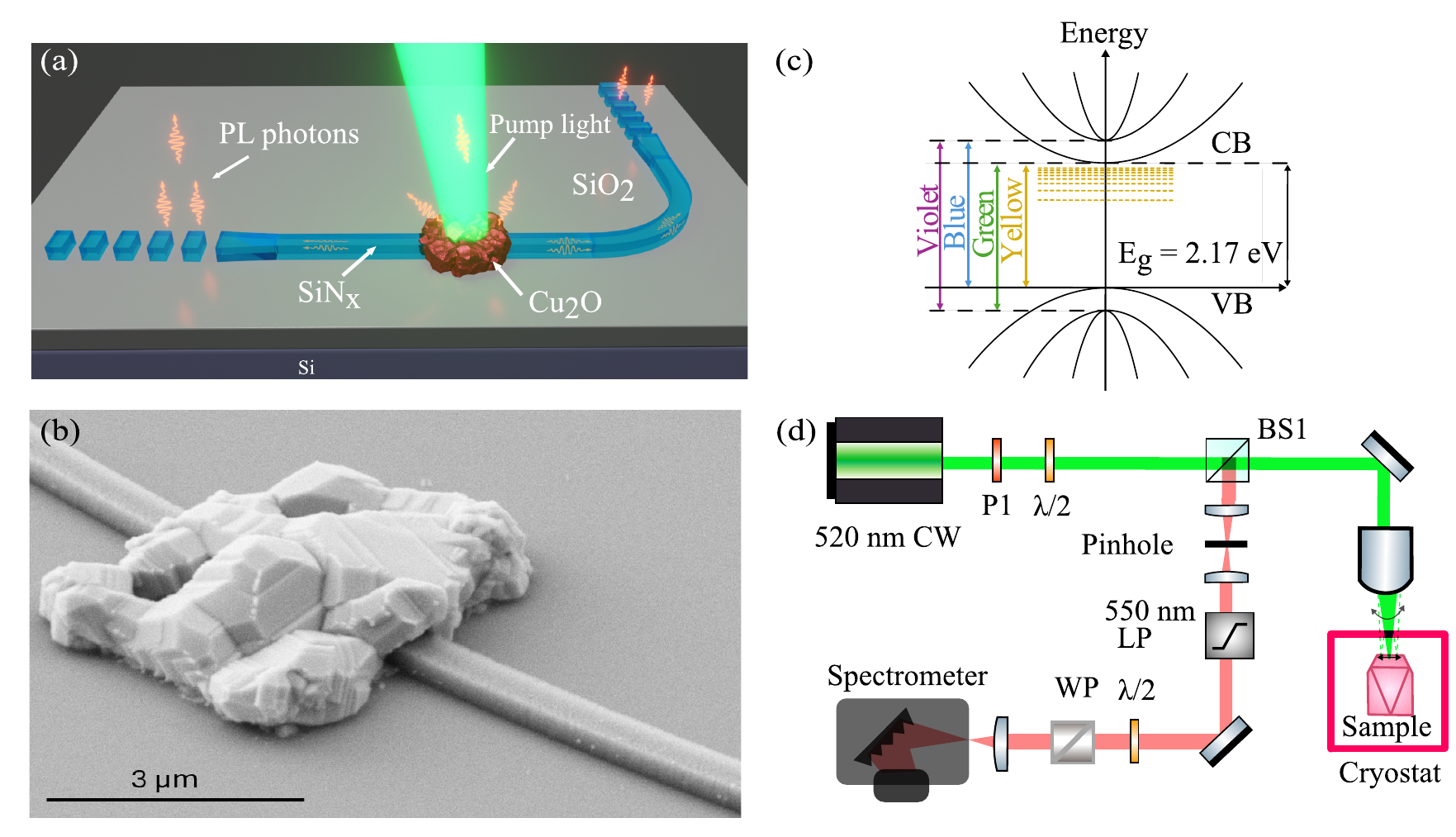}\\[0.5em]
    \caption{\textbf{Integrated \CuO{}-nanophotonic platform and excitation schemes.} (a) Schematics of the device, consisting of a \CuO{} microcrystal grown on top of a \SiN{} waveguide on a $\text{SiO}_2 /\text{Si}$ substrate with a $3.3\,\mu\mathrm{m}$-thick SiO$_2$ layer. The \CuO{} region is optically pumped from above, generating photoluminescence that can couple into the waveguide mode and be collected via the grating couplers at the end of the waveguide. (b) High-magnification SEM image of the \CuO{} microcrystal after growth. (c) Simplified band diagram of \CuO{} showing the valence band (VB), conduction band (CB), bandgap energy $E_g = 2.17$~eV, and the main excitonic series, yellow, green, blue, and violet. The horizontal yellow dashed lines represent Rydberg exciton energies within the bandgap. (d) Simplified optical setup schematics used for the excitation and collection, along with the polarizing optics used in the excitation and collection path. \textbf{P}, \textbf{BS}, and \textbf{WP} correspond to the linear polarizer, non-polarizing beamsplitter, and Wollaston prism, respectively. }
    \label{fig:device_and_experiment}
\end{figure}

\section{Results and Discussion}~\label{sec:results}

\subsection{A CMOS-compatible nanophotonic interface for \CuO{} excitons}~\label{subsec:cmos}
The platform is shown in Fig.~\ref{fig:device_and_experiment} and consists of a \CuO{} microcrystal (often referred to as a \CuO{} island in the text) grown on top of a \SiN{} ridge waveguide on a SiO$_2$/Si substrate. The waveguide has two ports, each terminated by a grating coupler for out-of-the-plane access: port G1 is $L \approx$ 45~\um ~away from the \CuO{} island through a straight section, and port G2 is $L\approx$ 30\,\um~away through a 90$^\circ$-bend of radius $R=50$~\um. While the two orthogonal ports enable proper cross polarization in resonant excitations, in the present non-resonant study, they serve no distinct role. The island is pumped non-resonantly ($\lambda$ = 520~nm) from free space above (green big beam in Fig.~\ref{fig:device_and_experiment}(a), and its emission is read out at G1 (red wave packets in Fig.~\ref{fig:device_and_experiment}(a).

As shown in Fig.~\ref{fig:device_and_experiment}(c), \CuO{} has a direct bandgap of $E_g\sim$2.17~eV and supports several excitonic series -- often referred to as yellow, green, blue, and violet -- associated with its valence/conduction-band structure. The yellow series is the most relevant for Rydberg exciton physics because it hosts the high principal-quantum-number $n$ states with large wavefunctions, large dipole moments, and strong long-range interactions. Under nonresonant green excitation, carriers relax into excitonic states and recombine, producing photoluminescence in the red/yellow spectral range. At cryogenic temperatures, phonon broadening is reduced, enabling spectral features associated with the excitonic manifold to be resolved. 

The waveguide was designed to account for the challenges of measuring Rydberg excitons in \CuO{}. Excitonic emission is orders of magnitude weaker than the phonon-assisted 1$s$ emission, so any photoluminescence from the waveguide material itself is undesired. To avoid autofluorescence of the waveguide, a 200-nm-thick nitrogen-rich SiN$_x$ film ($n\approx1.70$ at $\lambda = 570$\,nm) deposited by high-density plasma chemical vapor deposition, rather than stoichiometric Si$_3$N$_4$ was used; the former shows markedly lower autofluorescence under green excitation.\cite{Senichev2021,SenichevACS2022,MartinPhotophysics2023}. Due to the lower refractive index contrast between the \SiN{} and the SiO$_2$ substrate ($\Delta n \approx 0.25$), achieving effective mode confinement required etching down an additional 160 nm into the SiO$_2$ to form a pedestal. Electromagnetic simulations (see Methods, \nameref{subsec:device-des}) were used to design the waveguide, with the calculated optical loss exhibiting a minimum near the operating wavelength $\lambda = 575$\,nm (cf. Fig.~\ref{fig:waveguide_design}). Optical access is provided by grating couplers rather than cleaved facets, so that devices can be anywhere on the chip and probed in a confocal microscope through the cryostat window.

The \CuO{} microcrystal region is defined lithographically on the finished photonic device. Waveguides, gratings, and
alignment marks are patterned in a first electron-beam lithography step and transferred by CHF$_3$/O$_2$
reactive-ion etching. A second, aligned lithography step defines a lift-off window at the waveguide interaction
region, where a 20-nm-thick Ti adhesion layer and a 500-nm-thick Cu film are deposited by evaporation.
Subsequent thermal oxidation at 850$^\circ$C in 750 mTorr O$_2$ converts the copper into a \CuO{}
microcrystal in place~\cite{DeLange2023, Barua2025}. The Ti adhesion layer may also undergo oxidation during this process, potentially modifying its optical properties. To reduce the potential optical influence of the Ti adhesion
layer, the waveguide pedestal is also etched, increasing direct contact between \CuO{} and the SiN$_x$ waveguide
sidewalls (see Methods, \nameref{subseb:device-fab}). Scanning electron microscopy (SEM) in
Fig.~\ref{fig:device_and_experiment}(b) shows the resulting faceted \CuO{} island, approximately 4~\um{} across and 500~nm tall, straddling the 600~nm-wide waveguide at the designed position.

\subsection{Waveguide coupled cryogenic exciton emission}~\label{subsec:WG coupling}
The sample was cooled down to cryogenic temperatures and probed in a scanning confocal microscope. The sample was excited with above bandgap excitation, and the collected light was spatially filtered by a confocal pinhole and then spectrally filtered to remove the pump. The signal was directed to a Single-Photon Avalanche Detector (SPAD) for generating a photoluminescence map and then redirected to the spectrometer for collecting the exciton emission spectra, as depicted in the simplified setup schematics in Fig.~\ref{fig:device_and_experiment}(d) (cf. Methods and Fig.~\ref{fig:optical_setup} of Supplementary Information for more details on the confocal microscopy and spectroscopy setup).  

\begin{figure}[h]
    \centering
    \safeincludegraphics[width=0.95\linewidth]{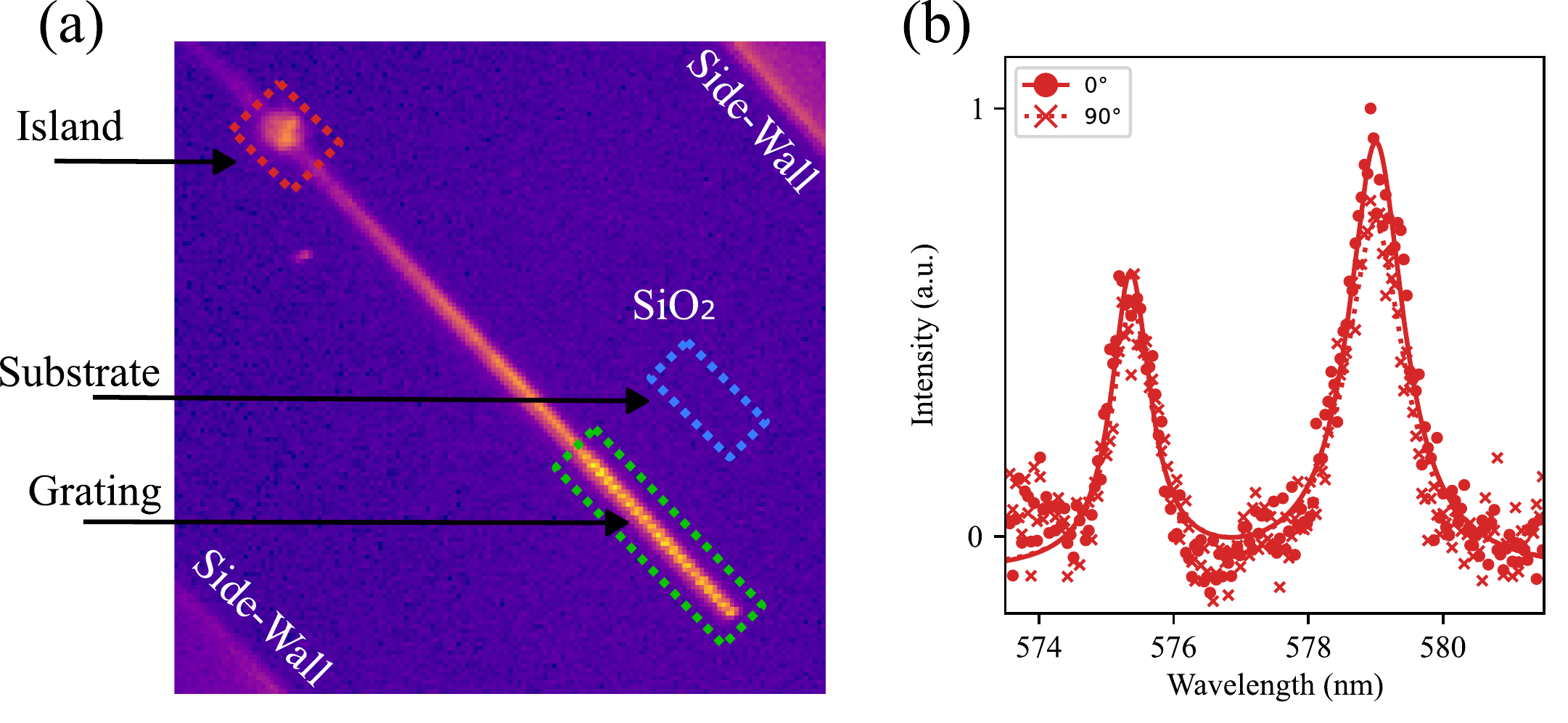}
    \caption{\textbf{Photoluminescence map of \CuO{}-waveguide device.} (a) A photoluminescence map of the \SiN{} waveguide showing the location of the grating coupler, the \CuO{} island, and the bare Si substrate next to the grating, highlighted with green, red, and blue dotted boxes, respectively. (b) Emission spectrum collected directly from the island after its excitation, clearly showing the emergence of low-lying $2p,3p$ levels from the yellow series. $ 0^\circ$ and $ 90^\circ$ correspond to the angle of the collection HWP before the WP, rotating the polarization of the signal by $90^\circ$. The spectrum shown is only the vertically polarized component of the signal, which shows the decrease in intensity when the signal is rotated by $90^\circ$.} 
    \label{fig:pl_waveguide}
\end{figure}

Figure~\ref{fig:pl_waveguide}(a) shows a confocal map of the long-pass-filtered photoluminescence emission, where the \CuO{} island, the \SiN{} waveguide, and the grating G1 can be clearly observed.  
Spectroscopy was performed in a local-excitation/remote-readout scheme: the collection objective was fixed on the grating G1, while the galvanometric scanner steered the excitation spot either onto the \CuO{} island, $L\approx$ 70~\um~away along the waveguide, or onto a bare substrate region adjacent to the island, as a control. With the island excited, the spectrum at G1, shown in Fig.~\ref{fig:full_series}, shows the two characteristic groups of features of the \CuO{} yellow series at liquid-helium temperature: a strong phonon-assisted 1$s$ orthoexciton band peaking near 614~nm (2.02~eV), accompanied by a weaker feature near 607~nm, and, roughly 3$\times$ weaker, two narrow lines at 578.8~nm and 575.0~nm (2.142 and 2.156~eV) corresponding to the 2$p$ and 3$p$ states of the yellow Rydberg series. 

When the excitation spot is moved to the adjacent substrate under otherwise identical conditions, none of these features appear at G1 (cf. Fig.~\ref{fig:full_series} of the Supplementary), and the residual signal is $\approx$ 50$\times$ smaller than the 2$p$ peak. The grating-collected emission therefore originates in the \CuO{} island and reaches G1 through the waveguide, rather than substrate fluorescence or direct optical scatter into the collection path. 

We fit the yellow series raw data with a double Lorentzian fit and extract the energy and full width half maximum (FWHM) of the peak. To support the assignment of the resolved features, we compare the peaks with hydrogen-like energies of the excitonic yellow series as in Ref.~\cite{Kazimierczuk2014}. Our results closely match the peak energies of the corresponding yellow series excitons present in literature. Table~\ref{tab:exciton_parameters} shows the fitted values from the spectra collected at different locations.

\begin{table}[htbp]
    
    \centering
    \begin{tabular}{lccc}
        \hline
        Collection & Transition & Energy (eV) & FWHM (meV) \\
        \hline
        Island (I) & $2p$ & $2.14136 \pm 0.00009$ & $3.66 \pm 0.23$ \\
        Island (I) & $3p$ & $2.15485 \pm 0.00011$ & $2.57 \pm 0.18$ \\
        Grating G1 & $2p$ & $2.14200 \pm 0.00030$ & $3.36 \pm 0.28$ \\
        Grating G1 & $3p$ & $2.15562 \pm 0.00014$ & $2.14 \pm 0.18$ \\
        \hline
    \end{tabular}
    \caption{Mean transition energies and full widths at
    half maximum (FWHM) obtained from fits
    for island (I) and grating (G1) collection. Values are reported
    as mean $\pm$ standard deviation across the nine collection
    angles shown in Fig.~\ref{fig:3}(b).\label{tab:exciton_parameters}}
\end{table}

\subsection{Polarization Signature of the Waveguide Mode}~\label{subsec:pol}
To distinguish guided photoluminescence from free-space scattering, we analyzed the polarization dependence of the grating-collected signal. The \SiN{} waveguide and grating couplers are designed to preferentially support and outcouple a TE-like guided mode (cf. Supplementary Fig.~\ref{fig:waveguide_design}). Therefore, the light routed through the waveguide and collected
by the grating coupler should be
strongly linearly polarized along a fixed axis set by the grating, irrespective of the random
dipole orientations in the \CuO{} microcrystal. 
Emission collected directly from the island, by
contrast, should be only weakly polarized, with any residual anisotropy set by the crystal
facets and the rough island morphology. 

We tested this with a polarization-resolved detection scheme in which a half-wave plate (HWP)
in the collection path rotates the polarization of the collected light, and a Wollaston prism (WP)
separates the horizontal (H) and vertical (V) polarizations into two distinct regions on the spectrometer detector (cf. the Methods section). The excitation conditions were held fixed, and only the HWP on the collection-side was rotated (cf. Fig.~\ref{fig:device_and_experiment}(d) and Fig.~\ref{fig:optical_setup} for the setup schematics). 

We used this polarization-resolved detection scheme first to characterize the emission from the \CuO{} island itself. We excited the island and collected the photoluminescence emission from the island itself by centering our objective lens on top of the island. The collected signal was split into two channels collecting the H- and V-polarized spectra. Fig.~\ref{fig:pl_waveguide}(b) shows the ``yellow series'' excitons in the V-polarized component of the emission. The collection HWP was rotated to observe a change in the intensity of the signal. The emission from the \CuO{} island is largely unpolarized, which is supported by the fact that the linearly polarized component of the signal is not completely extinguished with a $90^\circ$ rotation; see Fig.~\ref{fig:pl_waveguide}(b).

\begin{figure}[t]
    \centering
    \safeincludegraphics[width=0.95\linewidth]{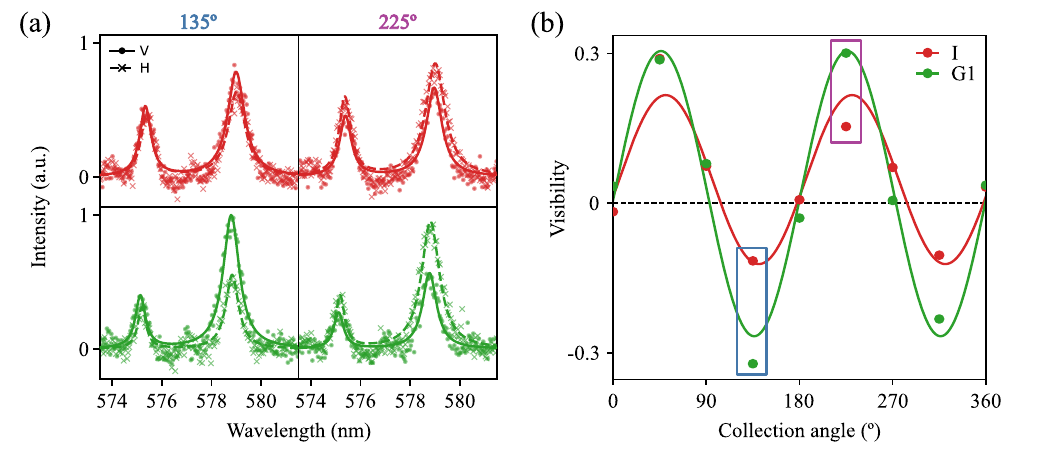}
    \caption{\textbf{Polarization dependence of the \CuO{} Photoluminescence for island and grating collection.} (a) Normalized fitted PL spectra measured at HWP angles of ($135^\circ$) and ($225^\circ$) for excitation on the island and collection from the island (red) or grating G1 (green). Solid and dashed curves correspond to the vertical (V) and horizontal (H) polarization components, respectively. (b) Visibility of the 2$p$ exciton emission as a function of the collection angle, as defined in Eq.~\ref{eq:visibility}, for island (I) and grating (G1) collection. }
    \label{fig:3}
\end{figure}

Figure~\ref{fig:3}(a) shows the H- and V-polarized spectra of the zero-phonon lines $2p,3p$ for two wave-plate angles, collected from the island (red) and from G1 (green). Figure~\ref{fig:3}(b) shows the normalized visibility ($V(\theta)$ in Eq.~\eqref{eq:visibility}) of the integrated 2$p$ emission
as a function of the HWP angle. Both signals oscillate with the expected period, confirming
linear polarization, but the modulation depth differs noticeably, as quantified through  visibility defined as

\begin{equation}\label{eq:visibility}
    V(\theta)=\frac{I_H(\theta)-I_V(\theta)}
    {I_H(\theta)+I_V(\theta)}\,.
\end{equation}

 This is calculated after correcting for the unequal detection efficiencies of the two WP regions (cf. Supplementary Fig.~\ref{fig:polarization_detection_bias}). The strongly enhanced visibility at G1 is the signature of a single guided mode acting as the collection channel: the waveguide and grating act as a polarizer on the otherwise weakly polarized excitonic emission. In the visibility measurements, there is residual asymmetry between nominally equivalent wave-plate angles, which reflects the asymmetrical behavior of the wave-plate due to artifacts like wedge angle. The total intensity fluctuates with an RMS value of $15\%$ over the entire angular range. The polarization contrast was hence normalized for each collection angle ($\theta$) (Eq.~\eqref{eq:visibility}) to account for these artifacts. This does not affect the comparison between island and grating collection, both of which were measured under identical conditions. 

 The resulting angular dependence in Fig.~\ref{fig:3}(b) was fitted to a sinusoidal modulation, $P(\theta)=P_0+A\cos(2\theta+\phi)$, where $P_0$ accounts for a possible polarization-channel offset, A is the modulation amplitude, and $\phi$ is the phase. The fitted modulation amplitude was used to estimate the polarization visibility, yielding ${V}_{I}=0.169\pm0.022$ for island collection and ${V}_{G1}=0.286\pm0.021$ for grating G1 collection. These results indicate a stronger polarization dependence for the emission collected through G1 compared with direct island collection.

\section{Conclusions and Outlook}~\label{sec:outlook}
We have demonstrated a monolithic, CMOS-compatible nanophotonic interface to \CuO{} excitons at cryogenic temperatures. \CuO{} microcrystals grown on a low-autofluorescence \SiN{} waveguide emit excitonic photoluminescence that is captured by the guided mode, routed along the circuit, and extracted at a grating coupler, where the phonon-assisted 1$s$ emission and the zero-phonon 2$p$ and 3$p$ lines of the yellow Rydberg series are resolved; spatial excitation controls and the enhanced linear polarization of the extracted light establish the guided--mode origin of the signal. Every element of waveguide QED with Rydberg excitons -- a Rydberg-grade crystal at a lithographically defined position, a low-background single-mode waveguide, scalable grating access, and a cryogenic measurement geometry -- is thus in place on a single chip.

The platform occupies a position that neither of its precursors does. Atomic waveguide QED obtains the guided-mode interface at the cost of ultrahigh vacuum, cooling and trapping, and Rydberg states that are fragile within a few hundred nanometers of a dielectric surface;~\cite{Vetsch2010, Kubler2010, Goban2014, Zhou2024PRX, Rajasree2020} here the Rydberg medium is a solid in direct contact with the waveguide, placed by lithography, coupled to the mode, and robust to thermal cycling. The length scales are favorable: the exciton diameter exceeds 1~\um~for $n \geq 15$, and the blockade radius reaches several microns~\cite{Kazimierczuk2014, Heckotter2021} larger than the guided-mode cross section, so that a single blockaded exciton could, in principle, gate the entire mode---the ``superatom'' regime that atomic systems reach only with optically dense ensembles.~\cite{Peyronel2012, Firstenberg2016} Relative to quantum dots, color centers, rare-earth ions, and 2D emitters on nanophotonic circuits,~\cite{Lodahl2015RMP, Sipahigil2016, Peyskens2019, Wan2020, Kindem2020, Parto2022} it trades single-emitter indistinguishability for an interaction-driven, cavity-free nonlinearity whose range is tunable with $n$, while retaining the wafer-scale growth, lithographic placement, and standard \SiN{} circuitry that made those platforms scalable.

The two-port, grating-accessed geometry is the natural setting for the follow-up studies on resonant spectroscopy of individual $np$ states through the guided mode, the interaction-induced nonlinearities predicted for Rydberg excitons in confined photonic modes,~\cite{Walther2018, WaltherPohl2020, Morin2022} and the probe of the exciton--exciton interaction directly.~\cite{Heckotter2021} The same fabrication process is compatible with microring and photonic crystal resonators, where exciton--photon coupling can be pushed toward the strong-coupling regime demonstrated in planar microcavities,~\cite{Orfanakis2022, Makhonin2024} and with the on-chip microwave-optical conversion.~\cite{Gallagher2022, Pritchett2024} Together with improvements in coupler efficiency and residual background that will bring higher-$n$ states within reach of the guided mode, these directions define solid-state Rydberg waveguide QED as an experimental program, of which the present work is the first step.

\clearpage
\newpage
\section{Methods}~\label{sec:methods}
\subsection{Device design}~\label{subsec:device-des}
Electromagnetic simulations were performed using the frequency-domain eigenmode solver implemented in Tidy3D (Flexcompute)\textcolor{red}{~\cite{}} to optimize the \SiN{} waveguide dimensions for supporting a single mode near the red and yellow excitonic spectral region of \CuO{}. The structure consisted of a 200~nm-thick \SiN{} ridge waveguide supported by a SiO$_2$ substrate with an approximately 160~nm-deep etched pedestal. The refractive
indices of nitrogen-rich \SiN{} and SiO$_2$ were taken
to be $n_{\mathrm{SiN}_x}=1.70$ and
$n_{\mathrm{SiO}_2}=1.45$, respectively, at an operating
wavelength of $\lambda=575$~nm.
The fundamental transverse-electric-like (TE-like)
guided mode was 
identified by the dominant in-plane electric-field
component perpendicular to the local propagation
direction. 
The simulated electric-field
distributions confirmed the confinement of the fundamental
guided mode predominantly within the SiN$_x$ core,
with reduced field penetration into the SiO$_2$
substrate due to the etched pedestal geometry.

\subsection{Device Fabrication}~\label{subseb:device-fab}
The photonic chip was fabricated using aligned multilayer electron-beam lithography to integrate \SiN{} waveguides with \CuO{} microstructures. To suppress autofluorescence under green excitation, a nitrogen-rich \SiN{} film with a reduced refractive index ($n \approx 1.7$) was used instead of stoichiometric Si$_3$N$_4$. The substrate was a commercial silicon-on-insulator wafer consisting of a 525 $\mu$m Si handle and a 3.3 $\mu$m SiO$_2$ layer that serves as the lower cladding ($n \approx 1.45$).
After solvent cleaning, a 200~nm-thick \SiN{} layer was deposited by high-density plasma chemical vapor deposition from silane and nitrogen precursors. This thickness was chosen to support the fundamental TE-like guided mode near the \CuO{} yellow exciton wavelength ($\lambda \sim 575$ nm). No post-deposition annealing was performed to avoid the formation of intrinsic SiN single-photon emitters reported in annealed nitrogen-rich films.~\cite{Senichev2021}

Waveguides, grating couplers, and alignment markers were patterned in a first electron-beam lithography step using 350 nm ZEP530A-6 resist coated with a 20 nm Espacer 300Z conductive layer. Pattern transfer was performed by CHF$_3$/O$_2$ reactive-ion etching, fully etching the \SiN{} layer and extending approximately 160 nm into the underlying SiO$_2$. This created an SiO$_2$ pedestal beneath the waveguides to improve optical confinement despite the modest \SiN{}/SiO$_2$ index contrast. Alignment markers were etched simultaneously to a total depth of approximately 360~nm.

Cu microstructures were fabricated in a second aligned electron-beam lithography step using a 700 nm-thick bilayer ZEP530A-6 lift-off mask. After the deposition of 20 nm Ti and 500 nm Cu by electron-beam evaporation, lift-off defined the Cu structures. These were subsequently oxidized at 850$^\circ$C in an O$_2$ atmosphere at 750 mTorr to form \CuO{} microstructures integrated with the photonic circuit.~\cite{Barua2025, Steinhauer2020} The completed devices were characterized by optical microscopy and scanning electron microscopy. Representative images are shown in Fig.~\ref{fig:device_and_experiment}(b) and supplemental Fig.~\ref{fig:sem_device}.

\subsection{Cryogenic Optical Measurements}~\label{subsec:optical}
We used a 520 nm fiber-pigtailed Hubner Cobolt 06-MLD laser focused on a sample with a 50$\times$, 0.6 NA Nikon CFI60 TU Plan Epi ELWD objective, leading to a diffraction-limited laser spot with FWHM = 0.33~\um. The sample was cooled to 4.5 K using a Montana Instruments CryoAdvance 50 cryostat. The signal from the sample was separated from the excitation laser using a Thorlabs BS033 50:50 beamsplitter and filtered with a Thorlabs FEL0550 550 nm long-pass filter. For photoluminescence mapping, the signal was routed to a fiber-coupled Excelitas single-photon counting module (SPCM-AQRH-14-FC), while a Thorlabs GVS012 two-dimensional galvo system in a $4f$-configuration was used to scan the excitation beam. Sample images were recorded with a Basler ace2 camera (a2A1920-160ucPRO). Spectra were acquired with a Princeton Instruments SpectraPro HRS-300 spectrometer and a liquid-nitrogen-cooled PyLoN 400 CCD camera.

\subsection{Polarization-Resolved Measurements}~\label{subsec:pol}
A zero-order half-wave plate followed by a Wollaston prism was placed in the collection path immediately before the spectrometer entrance slit. The prism directs the horizontal and vertical polarization components onto two spatially-separated regions of the CCD, which were read out simultaneously. Rotating the HWP rotates the polarization of the collected light with respect to the prism axes while the excitation conditions remain fixed. The relative detection efficiency of the two channels was calibrated by sending a weakly linearly polarized 580~nm laser beam, reflected from the bare substrate, through the full collection path and recording the H and V signals as a function of wave-plate angle (cf. Fig.~\ref{fig:polarization_detection_bias}); all H and V spectra were divided by the corresponding calibrated efficiencies before computing the normalized difference and the visibility of Eq.~\eqref{eq:visibility}. 

\subsection{Spectral Analysis}~\label{subsec:analyses}
Spectra were baseline-corrected using the asymmetrically reweighted penalized least-squares (arPLS) algorithm,\cite{Baek2015} which iteratively estimates a smooth background while suppressing the influence of spectral peaks. Reference spectra acquired from the substrate and from a bare waveguide region were subtracted where indicated. Individual lines were fitted with Lorentzian profiles; the quoted linewidths are the fitted FWHM after deconvolution of the Gaussian instrument response. 

\begin{acknowledgement}
The authors thank Dr. Alexei S. Lagoutchev, Dr. Alexander Senichev, and Ruben Canora Alvarez for insightful discussions during the early stages of this project. This work was supported by the U.S. Department of Energy (DOE), Office of Basic Energy Sciences (BES), Division of Materials Sciences and Engineering, under Award No. DE-SC0025554, which supported the device fabrication and optical characterization. Ben Cerjan acknowledges support from the U.S. Department of Energy (DOE), Office of Science, Basic Energy Sciences (BES), through the Energy Frontier Research Center under Award No. DE-SC0025620, which supported the growth of \CuO{}. 
\end{acknowledgement}

\section*{Data Availability}
The experimental and processed data supporting the findings of this study are available on Zenodo at \url{https://doi.org/10.5281/zenodo.23955416}. The corresponding Python analysis code is available on GitHub at \url{https://github.com/Quantum-Nano-Photonics/Cu2O-Rydberg-WG}.

\section*{Competing Interests}
The authors declare no competing interests.

\clearpage
\newpage
\setcounter{figure}{0}
\renewcommand{\thefigure}{S\arabic{figure}}
\section*{Supplemental Figures}

\begin{figure}[h]
    \centering
    \safeincludegraphics[width=0.95\linewidth]{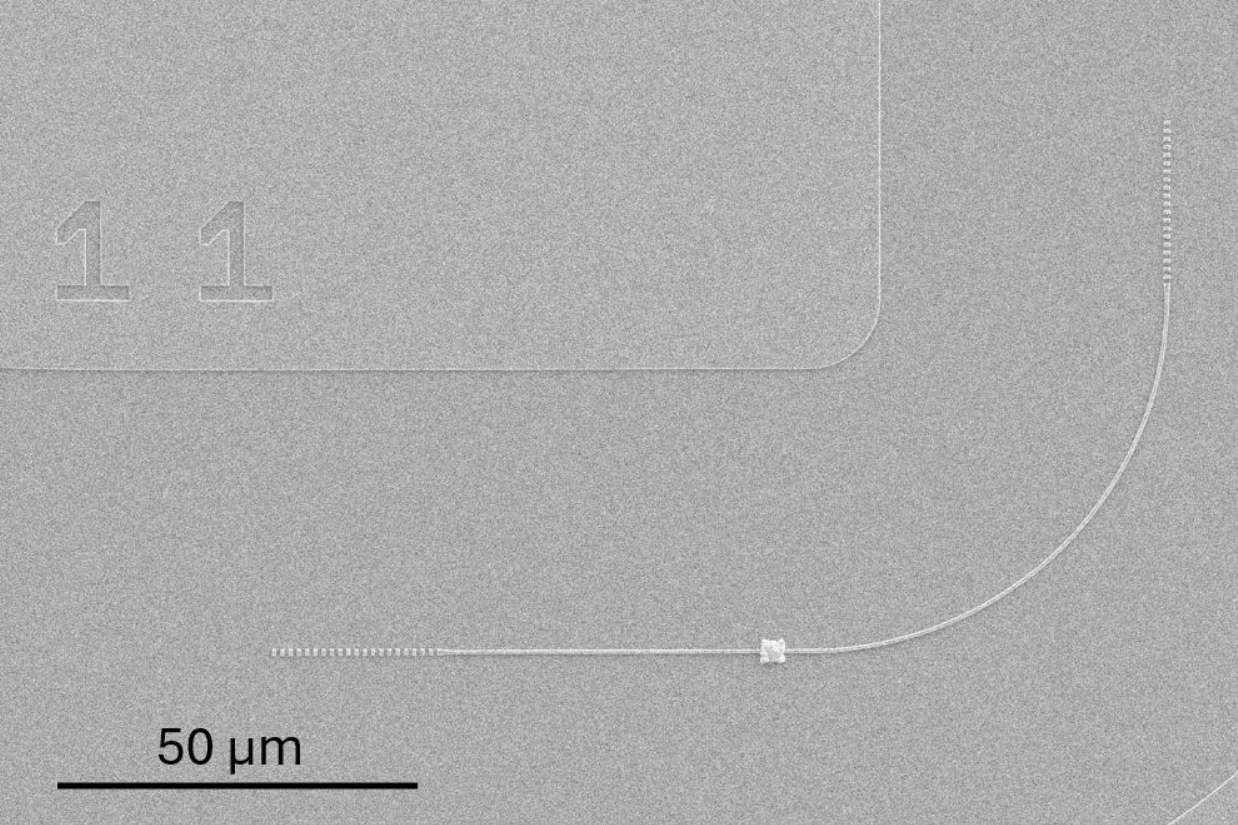}
    \caption{\textbf{Scanning electron microscopy images of the fabricated \SiN{}--\CuO{} integrated photonic device.} SEM image of the fabricated curved \SiN{} waveguide integrated with a \CuO{} microstructure positioned at the center of the interaction region.}
    \label{fig:sem_device}
\end{figure}

\begin{figure}[h]
    \centering
    \safeincludegraphics[width=0.95\linewidth]{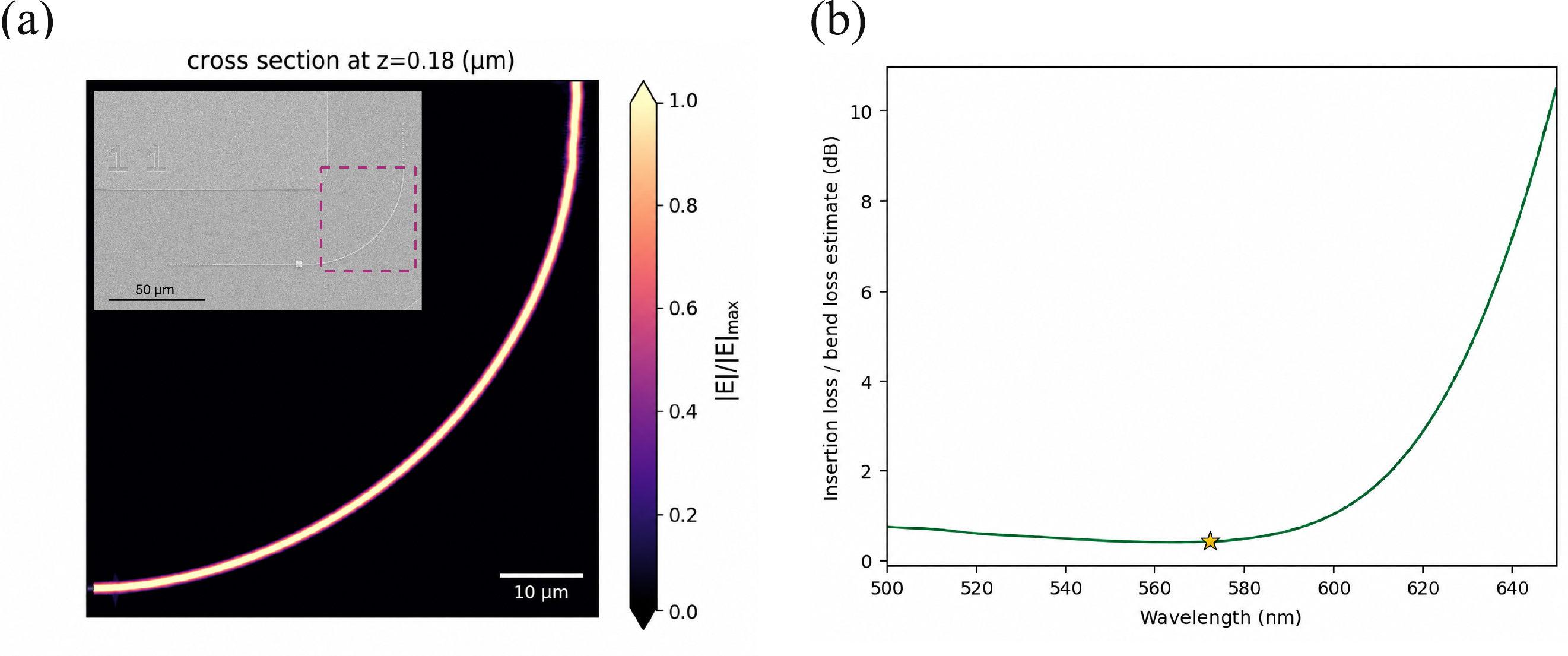}
    \caption{\textbf{Simulated \SiN{} waveguide design for \CuO{} excitonic photoluminescence.} (a) Simulated electric-field amplitude of the fundamental TE-like guided mode propagating through a 90$^\circ$ curved \SiN{} waveguide with bend radius $R = 50~\mu$m at $\lambda = 575$ nm. The \SiN{} core thickness is 200 nm, and the waveguide is supported by an approximately 160 nm SiO$_2$ pedestal. The inset shows an SEM image highlighting the part of the waveguide that is simulated. (b) Simulated wavelength-dependent bend-loss estimate for the optimized geometry, showing low-loss transport near the \CuO{} red/yellow excitonic spectral region.}
    \label{fig:waveguide_design}
\end{figure}

\begin{figure}[h]
    \centering
    \safeincludegraphics[width=0.95\linewidth]{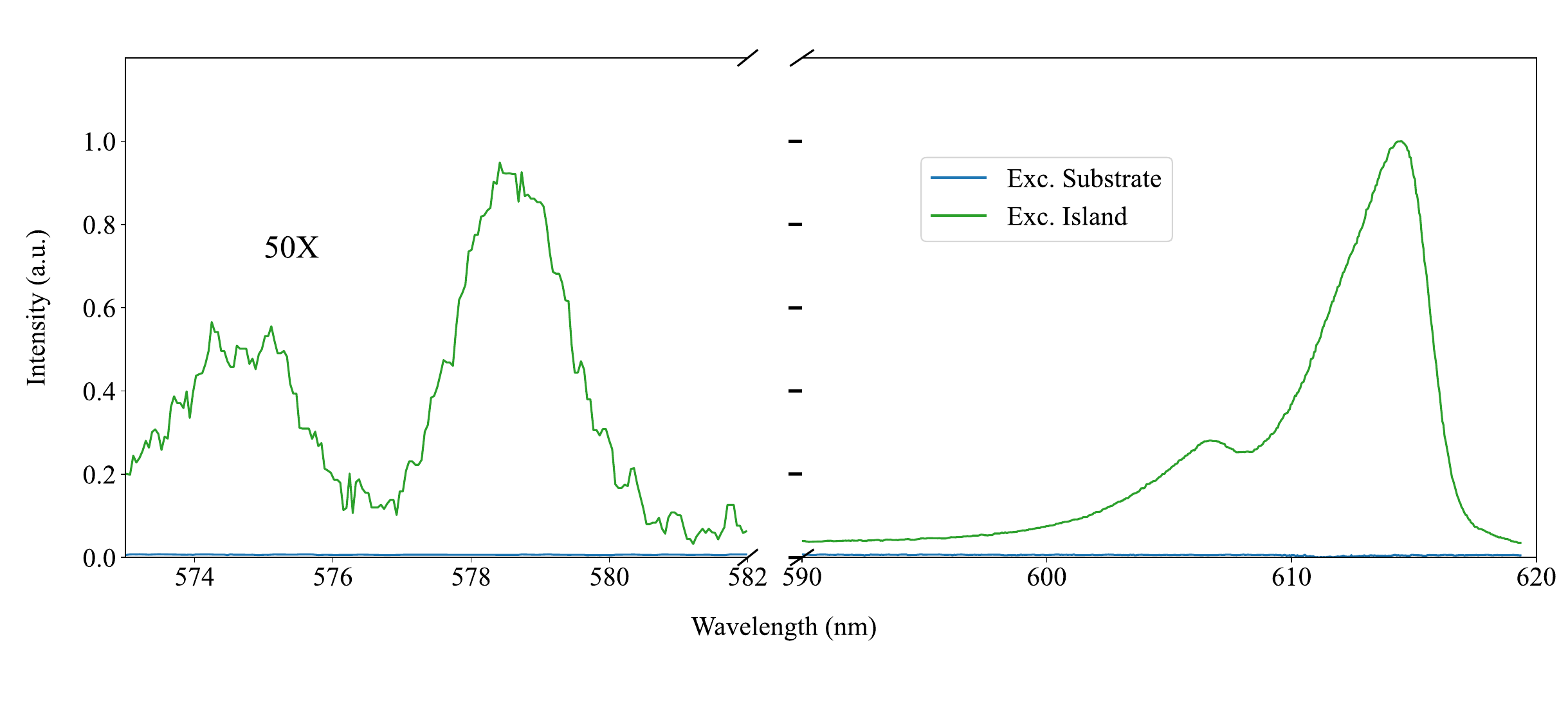}
    \caption{\textbf{1s and yellow series excitons in \CuO{} photoluminesence signal.} Spectral information corresponding to the ``phonon replica'' and ``yellow series” of the \CuO{} from the grating coupler signal after excitation at the \CuO{} island and substrate. The yellow series exciton peaks are scaled for clarity.}
    \label{fig:full_series}
\end{figure}

\begin{figure}[h]
    \centering
    \safeincludegraphics[width=0.95\linewidth]{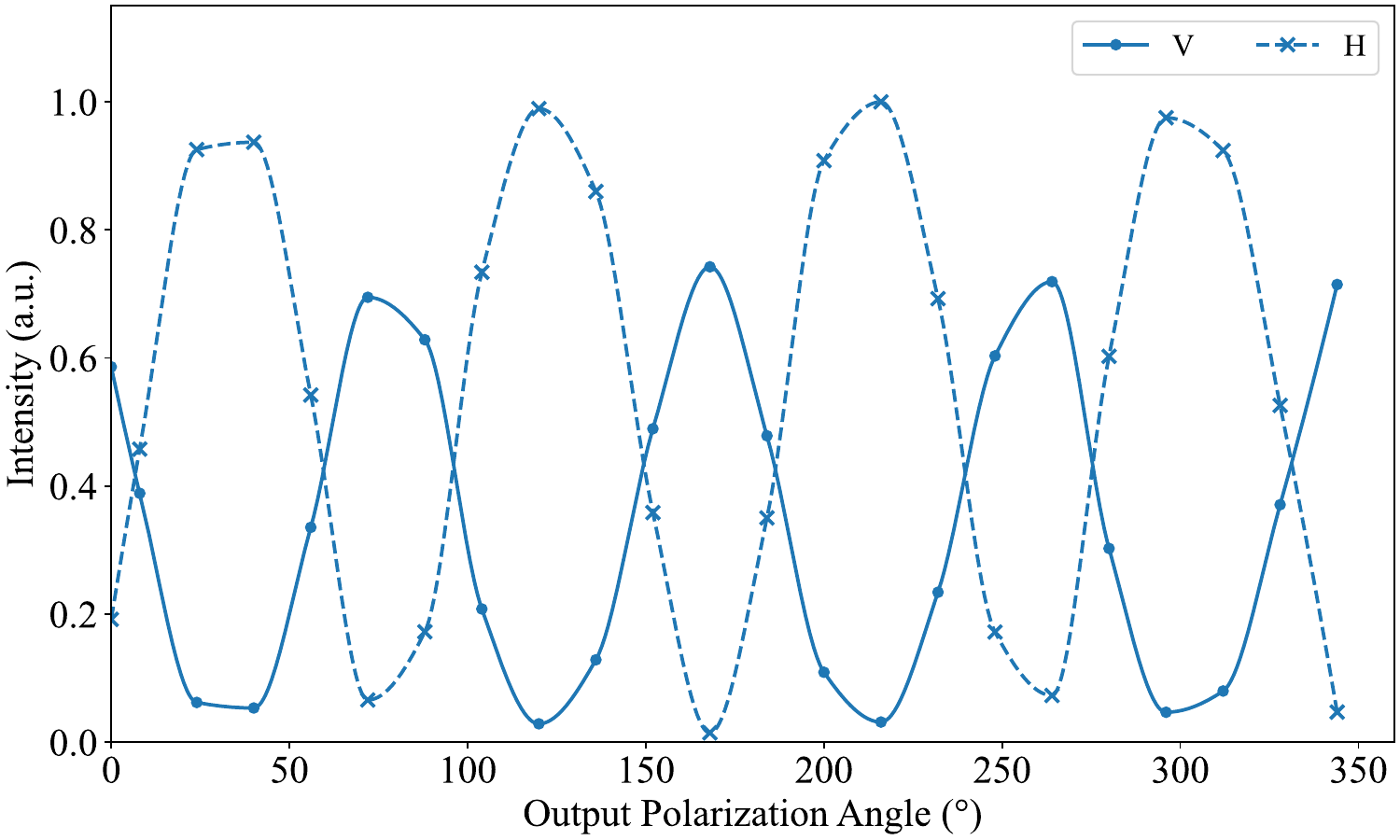}
    \caption{\textbf{Polarization Dependent Collection Bias.} A Wollaston prism is used to separate the horizontal (H) and vertical (V) components of the collected signal by directing it onto different spatial locations on the spectrometer CCD camera. The camera has different collection efficiency for different spatial locations, which was calibrated by sending a weak 580 nm laser, reflected off the silicon substrate, passing through all the collection optics, into the spectrometer. A half-wave plate was used to rotate the polarization of the collected signal and record the maximum and minimum signal collected at the spatial locations of V and H. The figure shows the oscillations in both polarization regions, clearly showing the difference in amplitude of oscillations. From this data, we can calculate the difference in the average signal across the two regions and characterize it as the collection bias across two orthogonal polarizations.}
    \label{fig:polarization_detection_bias}
\end{figure}

\begin{figure}[h]
    \centering
    \safeincludegraphics[width=0.95\linewidth]{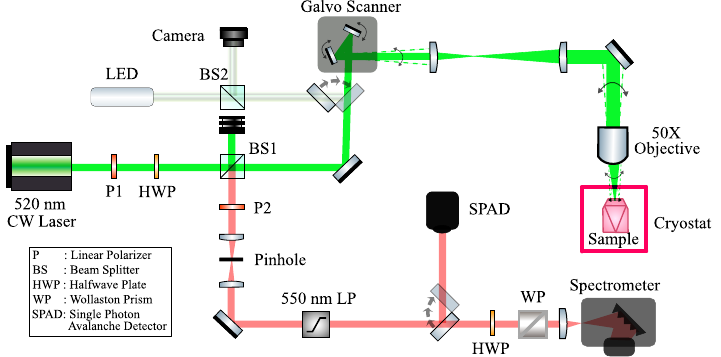}
    \caption{\textbf{Schematics of the experimental setup.} A 520 nm diode laser is used for linearly polarized excitation. P1 and P2 are linear polarizers, and BS1 and BS2 are unpolarized beam splitters. A galvo Scanner and SPAD are used to raster scan the sample to generate a PL map for the \CuO{} device. The signal from the sample follows the same excitation path until it is separated at BS1. It is spatially cleaned using a pinhole assembly and spectrally cleaned using a 550 nm long-pass filter. A Wollaston prism (WP) is used to separate the horizontal and vertical components of the collected emission.}
    \label{fig:optical_setup}
\end{figure}

\clearpage
\newpage
\bibliography{references}

\end{document}